# Embedding digital participatory budgeting within local government: motivations, strategies and barriers faced

Jonathan Davies

Computer Science, University of Warwick, Coventry, UK, jonathan.davies.1@warwick.ac.uk

Miguel Arana-Catania

School of Aerospace, Transport and Manufacturing, Cranfield University, UK, miguel.aranacatania@cranfield.ac.uk

Rob Procter

Computer Science, University of Warwick, Coventry, UK & Alan Turing Institute for Data Science and AI, rob.procter@warwick.ac.uk

The challenging task of embedding innovative participatory processes and technologies within local government often falls upon local council officers. Using qualitative data collection and analysis, we investigate the ongoing work of Scottish local councils seeking to run the process of participatory budgeting (PB) within their institution, the use of digital platforms to support this and the challenges faced. In doing so this paper draws on empirical material to support the growing discussion on the dynamics or forces behind embedding. Our analysis shows that formal agreement alone does not make the process a certainty. Local council officers must work as mediators in the transitional space between representative structures and new, innovative ways of working, unsettling the entrenched power dynamics. To do so they must be well trained and well resourced, including the ability to use digital platforms effectively as part of the process. This provides the necessary, accessible, transparent and deliberative space for participation.

## 1 INTRODUCTION

In an attempt to reconnect political institutions that are becoming increasingly detached from society, there has been a turn towards deepening citizen participation in local government processes. Such developments challenge representative practices and present deliberative and participatory methods as a way to strengthen democracy and enhance the legitimacy of public services.

Normative perspectives correctly portray citizen involvement [12] or the attitudes of political leaders [21] as critical to the process. Here, we shift attention away from these two groups towards another crucial issue for embedding participatory methods and the digital tools, the local government workforce. In doing so we explore their drivers and motivations, the processes and strategies they adopt, and the barriers and challenges they face.

This paper investigates the embedding of participatory budgeting (PB) and the use of digital PB within local councils across Scotland. PB is an innovative process that gives citizens the opportunity to participate directly in budgetary decision making, enabling local people to make collective decisions to address their priorities. Scotland has recently experienced widespread growth in PB, moving from a handful of highly localised community grant processes to it being embedded within the 32 local councils, or local authorities, that make up Scotland. This move to the mainstream is not simply a scaling up but represents a cultural change where decisions are made on the allocation of existing resources through both online and offline processes. This paper therefore seeks to answer the following questions: (i) what are the attitudes, the strategies adopted, and the barriers faced by local council officials when attempting to embed PB and, (ii) how can this workforce ensure that the mainstream process becomes fully connected to the local authority?

To explore this, we investigated the work of Scottish local councils via review of working group reports, meeting minutes, council webpages and qualitative data from 28 semi-structured interviews with local council officers. This provides much needed empirical material for an in-depth analysis of the embedding of participatory processes and their governance. We begin with a review of the growth of PB in Scotland, from experimentation to embedding, investigate what embedding entails and explore the role of digital PB in Scotland. We then present our research methods and analysis of the results, which reveal the challenges local council officers face in embedding the PB process successfully. Finally, we discuss our findings and future work.

## 2  RELATED WORK

There has been substantial research on the deepening of citizen participation in political decision making. Institutions or processes designed to do so have been classified as 'democratic innovations', a term initially defined by Smith [24] and further developed by Elstub and Escobar [13] who provide a comprehensive overview of this growing field.

PB, itself a democratic innovation, originated from Porto Alegre, Brazil in 1989. Early analysis on this process consequently focused on this city as a single case study. Both Abers [1] and Baiocchi [3] carried out qualitative research (namely interviews and ethnographic investigation) to address the mobilisation and participation of Porto Alegre residents in PB processes [1] and how PB provided space for necessary space for citizen participation through deliberation [3]. The subsequent diffusion of PB across the globe presented the opportunity to investigate multiple cases. Cabannes [7], for example, compares case studies from four cities in South America on the inclusion of young people in PB processes. As PB reached Scotland in the form of community led grant making, notable research from Harkins et al. [18], Escobar et al. [14] and O'Hagan et al. [22] centred on the evaluation of this early process and proposed mainstream PB as an area of future research.

By analysing multiple local authorities, each attempting to embed the PB processes, this research shifts attention away from singular [1, 3], exemplary [24] or experimental [6] case studies. In doing so, focus is given to the work of local council officials, rather than the well-researched areas of citizen participation in the process [1, 7] or the spaces of deliberation [3]. Finally, as Smith [25] has called for, by empirically analysing the use of digital PB as part of the process, this research revitalises the 'innovative' part of democratic innovations introducing a new, potentially disruptive form of engagement and tool to embed.

## 3 THE EXPERIMENTATION OF PB IN SCOTLAND

Initial growth of PB in Scotland was, in part, a result of grassroot action and demand for reform matched with top-down political and legislative support from government. This drive was at the centre of a national agenda of democratic innovation and community empowerment [14].

This early momentum provided the motivation for the Scottish Government to implement the national Community Choices programme. Between 2014-2018, £6.5 million was allocated to support and promote PB across Scotland through community led small grants. This was delivered in partnership with local councils (who often matched funding provided), community organisations and communities themselves [22]. This marked a shift away from processes primarily led by public agencies, to one where funding was targeted directly at citizens and citizen groups. This is referred to as first generation PB. By directly involving these groups in the decision making process, communities were given greater capacity to respond to local issues [14].

In 2017 the Scottish Government and Convention of Scottish Local Authorities (COSLA) agreed that at least 1% of local authority budgets would be subject to PB by the end of 2021 [9]. This represents over £100 million of funding, consisting of both revenue and capital expenditure. This move is commonly referred to as second generation PB or 'mainstreaming' PB and signifies the embedding of the process within council services as a way of working. It goes beyond the experimentation of allocating small grants from specific, pre-identified funding pots towards citizen participation in the allocation of existing resources across council services [22], increasing the local knowledge and opinions available to councils undertaking service planning [9].

Before exploring the role of digital PB within this process, it is first necessary to explore what embedding is.

## 4 EMBEDDING

Embedding participatory processes within an institution is a difficult task. In this section, we follow the definition of Bussu et al. [6] to understand and explore how these innovative processes often roll in "waves of experimentation and excitement" only to "break into disappointment" [6 p.1-2]. This break in momentum prevents the processes from moving beyond the periphery of policy making. By remaining at the margins, the processes fail to connect in full to institutions and the wider public sphere, limiting their legitimacy and success. This disconnect can often result in a failure to deliver any meaningful insights that may be produced. There is also concern that an un-embedded process does not fall in sync with existing political cycles but, instead, disrupts and displaces other forms of action [6].

By providing a much needed definition of embeddedness, Bussu et al. [6] begin to give reasons why these waves of experimentation break. They move away from formal thinking of embedding participatory processes as systematically coding formal rules and instead focus on broader, informal practices of transformation, adaptation and culture change. This shifts attention away from institutional design and gives more focus to the forces behind embedding, including the relationship between the process itself, citizens and the officials and practitioners implementing it. To address what embeddedness entails, Bussu et al. [6] outline three dimensions; temporal, spatial and practices. In doing so they provide a framework to guide future discussion and research.

1. Temporal: an essential aspect to embeddedness is the ability to repeat a process at regular intervals over time. This repetition may result from the process taking a permanent position in a political cycle [8] or simply from its informal habitual use [15]. Bussu et al. [6 p.5] suggest that the permanent structures and informal practices can merge, stating "iterative participatory structures can shift the informal practices of public agencies frontline practitioners and autonomous grassroots actors, fostering a sustainable participatory culture." It is at this alignment where embeddedness begins to take place.
2. Spatial: the spatial dimension can incorporate two facets: firstly, community access to, and influence on, spaces of decision making and second, how participation may link such spaces to wider civil society. We may divide decision making spaces into spaces within government and spaces within policy. The local level is viewed as the most conducive space to embed participation through practical problem solving on issues close to citizens [16]. For policy space, Fagotto and Fung [15] argue that a process is embedded when it is 'encompassing', when it spans several policy issues, rather than one single issue. Finally, Bussu et al. [6] shift focus away from political and policy-making institutions to embedding participation in relation to civil society, the act of a community 'anchoring' a process and the critical energy of civil society.
3. Practices: focus has often centred on the formal rules of embedding, which stem from, and are framed around, specific institutional designs or the right of individuals and communities to participate [6]. For example, the formal agreement between the Scottish Government and COSLA that at least 1% of local authority budgets would be subject to PB. Although this formalising or ordering is important, Bussu et al. [6] place emphasis on the informal, actor-centric account of practices of embedding. This gives focus to the actors who create and weave new relationships between organisations, institutions and communities, and help to challenge the existing structures in place.

**5 EMBEDDING DIGITAL PB**

An important part of the mainstreaming of PB in Scotland is the exploration of digital PB to further enhance democratic processes and the effectiveness of local government. When investigating the move from experimentation to embedding it is therefore beneficial to examine the move from the experimentation of suitable digital tools to the embedding of one tool to support the mainstream PB process. It should be noted that to avoid any form of technological determinism, the focus here is not on the tools alone, but how they provide support, are used, and work within the wider assemblage.

In 2015, the Democratic Society was commissioned by the Scottish Government to produce a report on the use of digital tools to strengthen the PB process within local councils. The organisation worked with 11 local councils to support the adoption of digital elements in PB processes. The report recommended that digital tools are an important support for offline engagement and can allow for increased and wider participation, flexibility in engagement and enhanced connectivity to local networks. The report also recognised a number of potential challenges, including the creation of a digital divide, the amplification of existing offline voices, the polarisation and fragmentation of communities, and the balance between security and accessibility [11].

One such digital tool for citizen participation is the collaborative free software, Consul. The project was inspired by the Icelandic platform 'My Neighbourhood' and has been used by 35 countries and 135 institutions across the world including, for example, the city of Madrid, with €100 million decided by citizens each year, and Porto Alegre, the birthplace of PB. A range of NLP-based tools were recently developed and integrated into the Consul platform [2]. These tools give the platform the potential to facilitate more effective collaboration between users by tackling the problem of 'information overload' – when the amount of information available is difficult for users to process, or where users find it difficult to interact with each other. In 2020, COSLA approached us about deploying the digital platform Consul across their future mainstream PB programmes. This presented the opportunity to not only investigate the challenges of embedding PB across local councils in Scotland, but also how an innovative digital platform may support this.

## 6 METHODOLOGY

For an in-depth understanding on how the PB process is embedded and supported by digital PB within a local council, and to give voice to the attitudes and approaches of local government officials, we engaged in qualitative data collection and analysis. This includes a combination of secondary research and semi-structured interviews. Below we present each stage of the implementation of our research method, which as Mackieson et al. [20 p.1] states can effectively increase "rigour and transparency", thereby reducing potential bias, something qualitative analysis is often criticised for.

The initial stage was a review of existing literature, essential for deepening insight and awareness as well as developing our research questions. As analysis is an iterative process, and this research is ongoing, this review is also ongoing. This turn and return to new or existing literature facilitates a "creative interplay among the process of data collection, literature review, and researcher introspection" [23 p. 226].

To move beyond any a priori assumptions that may emerge from a literature review, secondary research was carried out. This provided contextual information on the state of PB in Scotland, the current state of PB within each local council, and information on the local councils themselves. Research included assessment of local council websites, committee meeting minutes, council reports, and attendance of events organised by the Scottish Community Development Centre and the PB Scotland Network.

Secondary research was complemented by a longitudinal study using a cohort approach. This involves repeated observations, in this case semi-structured interviews of local council officers, over a period of time. This form of data collection is key to forming causal interpretations and documenting changes in attitudes that emerge over time, something not possible with a more static cross-sectional study [10]. Local council officers were invited to interview so we might understand their attitudes and approaches to the PB process. Interviews were transcribed, uploaded to the software NVivo 12, and analysed using the process of applied thematic analysis (ATA).

### 6.1 Applied thematic analysis

As with qualitative data collection, there are a number of ways to approach qualitative data analysis. Rather than give focus to just one, "good data analysis… combines appropriate elements and techniques from across traditions and epistemological perspectives" [17 p.3]. As such, this paper uses the process of applied thematic analysis on the transcribed interviews. This is a type of inductive analysis that can involve, or sit between, multiple analytic techniques [17, 19], emulating parts of grounded theory, ethnography, phenomenology, and other qualitative methodologies [19]. ATA itself involves systematically identifying, organising and analysing repeated patterns of meaning, or themes, across the data set [4, 5]. By focusing on commonalities and differences across the data set, Braun and Clarke [5] highlight how this method allows one to make sense of common or shared meaning and is less suited for examining experiences from a single data set.

This is exploratory, content driven, inductive analysis. The patterns are not predetermined but instead derived from the data and supported by it. These themes are not just used to classify or label but also "reframe, reinterpret, and/or connect elements of the data" [19 p.3]. Thus, the method describes the data but also allows for interpretation. This provides structure but also encourages the practice of reflexivity [20].

After analysing multiple interviews, we were then able to carry out a cross-case comparison, identifying what codes are shared with each case and thus what each case has in common. This is important as it helps to distinguish between information or accounts which are relevant to all local councils and aspects of the experience that are exclusive to certain councils.

## 7 ANALYSIS

After exploring the initial motivations for PB in Scotland we present the results of our analysis in this section, exploring the local council attitudes towards the process, the strategies or approach they adopted, the barriers and challenges they face when attempting to embed PB and the digital tools associated with it. These arguments emerged from the analysis of the transcribed interviews. The interviews included questions on (1) current and past PB processes run (2) opportunities and challenges experienced as a result of these processes (3) perceived opportunities and challenges which come from embedding PB as a mainstream process, and (4) if use of the digital tools mitigate these challenges.

We identified two clear central themes. Firstly, the advantages and challenges faced by the local council as a result of how they approached first generation PB or community led grant making. Secondly, attitudes and approaches to mainstream budgeting.

### 7.1 Community led grant making

We can divide a council's approach to first generation PB into three types: a fully in-person approach, a fully online approach, and a blended approach of in-person events and online processes. Those interviewed discussed the advantages and barriers or challenges faced using each type.

It was found that local councils experienced a sense of community cohesion and connectedness when running in-person events. A number of councils noted that community groups and individuals whose ideas were not successful on the day of the event were still able to connect with other local council services and community partnerships to help implement their proposals, often through the identification of additional funding. Concern was raised that this experience would be very difficult to replicate online.

> "In previous years they've been almost a little bit of a celebration of all the community groups coming together, which is why the teams wanted to specifically run in-person events." (interview 5, Strategic Planning Manager)
>
> "… that kind of community buy in, that cohesion, the connections that can be made that just don't happen the same when you're on an online platform." (interview 2, Service Manager, Communities)

Despite this feeling of cohesion, there were a number of notable challenges experienced as a result of an in-person approach. Firstly, officials often received feedback from grant applicants who found it difficult to present their idea to the public using either a marketplace or presentation format. Secondly, concern was raised that the processes lacked public deliberation or discussion. Either members of the public would attend the event only to vote and leave without any other input or officials noted a lack of discussion during the idea approval stage. Thirdly, officials experienced a high workload when running an in-person process (for example, through the provision of voting material or the manual counting of votes). Finally, there was concern that the success of the approach depended heavily on members of the public attending the event.

> "… the feedback from individuals was it was quite stressful, having to go up and do that presentation." (interview 2, Service Manager, Communities)
>
> "… they had a sort of bidding project process and they had a small group from the community in conjunction with the partnership that just sifted out all the projects. They didn't have much discussion, they just said, yep, we think that's a good project, yes or no. And they sort of determined which ones went up." (interview 6, Strategic Planning Manager)
>
> "… it was through winter as well so you can't get people down to the town hall." (interview 7, Community Engagement Officer)

The majority of councils who had run an online event were optimistic about the process. The interviews revealed that many local councils experienced an increase in participation when the process was held online. This is supported with secondary evidence showing a greater number of ideas generated and votes cast online.

> "Online is the way forward as you're getting to that online audience, you know the digital people like myself who've sat in the home and want to just click buttons and don't want to go to the town hall on a Saturday afternoon thinking I better get home cleaning the house." (interview 7, Community Engagement Officer)

Despite evidence of greater participation, officials were acutely aware that running a fully online process may exacerbate issues of digital exclusion. They feared that communities with poor digital access or those who lacked digital skills may not be able to, or want to, participate online.

> "we had trialled that just as an online only and people really struggled. We didn't take on board the deficit and people's digital, you know, skills. So, some people really struggled and we were having to logon for them and talk them through it. So, it took a lot more, much more resource intensive." (interview 9, Natural Environment Officer).
>
> "I think I would say that one of the main challenges has been getting rural communities to participate. And vulnerable families." (interview 14, Service Manager, Communities)

Finally, a blended approach allowed for the mitigation of the challenges experienced during a fully online or fully in-person process. However, two distinct tensions emerged. Firstly, how to select the correct timescales of online and in-person events. For example, if the in-person event lasts for just one day, how long should the online process run for? Secondly, how should the voting online and in-person be weighted? One local authority struggled with this using an earlier version of the Consul platform:

> "They had an in person voting event and they kind of weighted the in person voting so that your first choice got three points, your second got two and one. So almost like if you turned up in person your votes had more weight. We would have done that equivalent on Consul, but Consul wasn't able to do that kind of weighted voting." (interview 4, Consultant, Community Development)

### 7.2 Mainstream PB

The majority of local authorities interviewed had not yet launched, or were in the process of launching, their mainstream PB program. Despite this, many were still able to provide opinions and discuss their intended method, motivations and concerns. We can divide these emerging themes into attitudes towards the turn to mainstream PB and the approach a local authority is taking.

*7.2.1 Attitudes to mainstream PB*

In regard to attitudes towards mainstream PB, discussion focused heavily on a cultural change, in particular, a new way of working for local authority officials. This includes opinions on where to situate the PB process within the local authority, including where 1% of the council's budget will be taken from (ownership of the 1% agreement), and which department will own and manage the digital platform (ownership of the platform). Many local authority officials noted that they were now seen as the spokesperson or ambassador for mainstreaming PB and the go-to person for questions regarding the digital platform.

> "We really need to get that commitment from our corporate management team and then approved, fully, through cabinet…we really need to get that buy in and commitment right across the services." (interview 14, Service Manager, Communities)
> "If they want to use Consul, they're kind of coming to me, probably this should be going more to the web team. I need to make the system more mainstream… what is our communities' digital offering and where does Consul sit within that?" (interview 4, Consultant, Community Development)

The interviews also revealed that the majority of councils experienced early resistance to the mainstream PB process from elected council members, presenting a political challenge.

> "Elected members really struggle with the whole concept of handing power to people. And are very, very reluctant to do anything near mainstream." (interview 5, Strategic Planning Manager)
> "That was one of the saddest things, people tearing the process apart just to score points off each other. It was really unfortunate because it's such a special process." (interview 10, Community Partnership Manager)

One final issue discussed was the view that the mainstream process will increase transparency of the services councils provide. By opening up discussion around the council budget, citizens were able to recognise the services provided and understand where money is spent. It was hoped that this transparency would also work the other way, where the staff running the process will feel more empowered with the knowledge that the work they were doing had been decided upon directly by the community.

> "it's that sort of recognition, I guess. It's beyond just simply deciding on a budget. For example, it's releasing this transparency and sort of releasing the information or the knowledge of what the Councils actually do." (interview 4, Consultant, Community Development)

*7.2.2 Approach to mainstream PB*

Mainstreaming PB requires finding solutions to the problems identified through process design. Although the focus of this paper is on the dynamics of embedding rather than institutional design, the forces and drive behind the design can only be understood by exploring the approach or rules adopted. Two key challenges emerged from this discussion, the approach to the operational challenges, and the task of managing legitimacy.

One major operational concern, stemming from the challenges of the online first generation PB, was access to the process - how local authority officials can reach and support marginalised groups when the process is run online. This is particularly important for the mainstream process, which involves opening up a dialogue with community members on redesigning or reconfiguring council services. Many councils have, or plan to have, measures in place to support marginalised groups, such as posting out voting forms (to then enter in manually), training library staff to help visitors vote, or in-person visits to groups or community centres with electronic tablets. However, concern was again raised of the workload this creates and that these methods would not provide the sought-after community cohesion experienced at in-person events.

> "it's shown that that postal bit works in some ways, but you know, when it comes to actually bringing a community together and making the connections between groups and organisations, and individuals who are maybe just new to the local area, it's a great opportunity to meet new people, find out what's going on in the area as well." (interview 14, Service Manager, Communities)

A further concern relating to access to the process was the registration method for the digital platform. Councils were often caught between wanting to have an open registration method to allow for individuals to participate and vote without requiring an account, but also wanting a secure registration method to prevent members of the public from voting multiple times.

> "Setting it up so the public could access it with their email address and their post code and things like that, that always seemed like a bit of an issue as to how that data, if that data was ever going to be kept. You're always going to wonder what the contract is." (interview 9, Natural Environment Officer)

Each local authority was aware of the need to measure, evaluate and report on the impact of the process. Councils were able to capture quantitative data, such as the number of participants or the number of votes, yet were conscious of the challenge of capturing important wider impacts, including assessing the difference a project has made within a community:

> "Our reporting has tended to be fairly basic in terms of this is what we funded, this is the number of people who participated." (interview 19, Community Planning & Development Manager)

And assessing the increase in transparency:

> "Again, it's that sort of recognition, I guess. It's beyond just simply deciding on a budget. For example, it's releasing this transparency and sort of releasing the information or the knowledge of what the Councils actually do. Which is difficult to measure…it's hard to be able to come back to say people have become more aware of what we do now." (interview 4, Consultant, Community Development)

Finally, the challenge of how to ensure the process is legitimate was raised. Officials were aware that robust mainstream PB is the effective allocation of public resources to areas most in need and not, for example, used to cut public spending or on projects deemed unnecessary. As council budgets are often set to strict parameters there was concern that members of the public would only be voting on pre-set choices with little input. Without discussion or deliberation and equitable voting, the process runs the risk of being tokenistic.

> "The only gripe that we've had thus far is because we're saying, 'you have £50,000 to upgrade this park', some people have come back and said that's not PB, when you tell me what we have to spend money on.

There's going to be a bit of an element of that with mainstreaming, I guess because people are so used to having quite loose criteria for small grants." (interview 8, Principal Accountant)

"Voting isn't fair, you know, if you got these big, well-established groups that have been on the go for years or semi-professional and know how to raise money to help get the vote out against the wee community groups… the big organisations… they hoover up all the funds." (interview 12, Community Planning Coordinator)

## 8 DISCUSSION

Having explored what embedding PB entails, the challenge is to explore how the local council officials ensure that the mainstream process becomes fully connected to the authority and the wider civil society rather than a process which sits disconnected from broader local governance as a scaled-up grant making programme. This involves not only a transformation of the local council's institutional design but investigation into the informal dynamics of embedding.

Temporally, the 1% agreement between the Scottish Government and COSLA marks a turn from ad hoc PB processes to a formal practice repeated over time, running in sync with the overall budgeting cycle for each local authority. Yet, although formalised through agreement, it is possible for the process to fail to become embedded within the institution [6]. To avoid this requires two actions; (1) the need to understand and encourage informal habitual use of PB within the local authority itself and (2) an understanding of how the PB process will work in conjunction with the temporal restrictions of the community. Firstly, local authority habitual use of the digital tool and of the process will emerge from better knowledge of the process and of the digital platform. Therefore, PB and digital PB must be supported by suitably trained teams. Any new internal ownership of the PB process and of the digital platform must be accompanied with sufficient training of both. Secondly, citizen participation will be greatest when any time constraints are removed. Interviews revealed that a blended approach, consisting of an online process and in-person events, allows for greater participation. Ensuring both (1) and (2) sees the alignment of informal practices and permanent structures allowing for embeddedness to begin to take place.

Concern was raised that to achieve the above, and to ensure that the online and offline processes work in sync with one another, increases workload. A trained team must therefore be well-resourced. This includes the provision of a well-designed digital platform which, when used effectively, allows for greater participation in the process [2]. This provides capacity to minimise barriers to participation and support inclusion, which underpins the legitimacy and effectiveness of PB. The legitimacy and effectiveness of the process can be successfully evaluated and assessed by a knowledgeable and well-resourced team. Demonstrating a successful process and a successful digital tool then allows for it to work in conjunction with other political cycles, such as local elections, that may threaten to unembed PB if its effectiveness cannot be proven.

Spatially, PB is most effective when communities have access to a decision space where they may actively influence and reshape the delivery of local council services. This is a process that is embedded at the local government level [15] and spans several policy issues rather than just one [16]. Interviews confirmed that community access to this level was considered a challenge with measures put in place by officials to support and allow for equal participation. Officials were also aware that the PB must not be seen as tokenistic and instead achieve community empowerment and public sector reform by "allocating resources to the areas of priority need" [9 p. 1]. This is supported by secondary research into the existing platform data, which shows that the ideas generated by community members and community groups tend to match closely with council priorities. To drive this fundamental shift in public service provision requires greater deliberation and discussion on ideas that match these priorities and equitable voting to ensure these ideas are successful.

Finally, despite an agreement to allocate 1% of local authority budgets by PB, interviews revealed that when embedding mainstream PB, there is a lack of established formal space for the PB process. As a result, officials are not only required to engage with local people but also act as mediators and negotiators between departments (during the internal cultural change of a new way of working), and broker new relationships between the existing representative system and the new, often radical, form of participation (a political challenge). They are faced with the task of challenging tradition and unsettling established power dynamics. It is here we see the 'practices' dimension of a commitment by officials to not only formalise the process, but create and recreate relationships between community members, local authority departments, elected representatives, and the digital tools.

## 9 CONCLUSION AND FUTURE WORK

In this paper we explored the barriers and challenges faced by local council officers when embedding PB in local government and the use of a digital platform to support this. In doing so we provide much needed empirical material to support discussion on embedding participatory processes.

Despite the introduction of mainstream PB through the 1% target, we found that a formal agreement alone does not make the process a certainty. Instead, the action of embedding is something that requires constant, often overlooked, essential work from local council officers. Included in this is working as a mediator or negotiator in the transitional space between the traditional representative structures and new, innovative ways of working. To be effective, officers responsible for embedding PB must be well trained and well resourced. This allows for the habitual use of PB, the power to pass on the process to other departments, the ability to put measures in place to support equal participation, and capacity for assessment and evaluation.

Importantly, officers must be trained to use a digital platform such as Consul effectively, alongside offline processes. The platform then acts as a resource, removing the barrier to inclusiveness often seen in face-to-face processes. Its effective use provides the necessary, accessible, transparent, and deliberative space for participation and co-production to occur [2, 10].

As these processes become embedded and start to run in sync with council budgeting cycles, future research will include deeper analysis into data from the process: ideas generated, selected for ballot, and voted on and then compared against local council priorities. This will determine if resources have indeed been equitably allocated to the areas of priority need and if the process design is sustainable and effective. Doing so provides the empirical evidence required to evaluate the wider social and democratic goods generated in the longer term, once embeddedness has taken place.

**ACKNOWLEDGMENTS**

This research has been supported by the Alan Turing Institute for Data Science and AI (grant no. EP/N510129/1) and by the EPSRC Impact Acceleration Fund (grant no. G.CSAA.0705).

**REFERENCES**
[1] Rebecca Abers. 1998. From Clientelism to Cooperation: Local Government, Participatory Policy and Civic Organising in Porto Alegre, Brazil. Politics & Society, 26(4): 511-537.
[2] Miguel Arana-Catania, Felix-Anselm van Lier, Rob Procter, Nataliya Tkachenko, Yulan He, Arkaitz Zubiaga and Maria Liakata. 2021. Citizen participation and machine learning for a better democracy. Digital Government: Research and Practice, 2(3): 1-22.

[3] Gianpaolo Baiocchi. 2003. Emergent Public Spheres: Talking Politics in Participatory Governance. American Sociological Review, 68 (1): 52-74.

[4] Virginia Braun and Victoria Clarke. 2006. Using thematic analysis in psychology. Qualitative Research in Psychology, 3(2): 77-101.

[5] Virginia Braun and Victoria Clarke. 2012. Thematic analysis. In H. Cooper, P. M. Camic, D. L. Long, A. T. Panter, D. Rindskopf, & K. J. Sher (Eds.), APA handbook of research methods in psychology, Vol. 2. Research designs: Quantitative, qualitative, neuropsychological, and biological (pp. 57–71). American Psychological Association.

[6] Sonia Bussu, Adrian Bua, Rikki Dean and Graham Smith. 2022. Embedding participatory governance. Critical Policy Studies. https://doi.org/10.1080/19460171.2022.2053179

[7] Yves Cabannes. 2005. Children and Young People Build Participatory Democracy in Latin American Cities. Children, Youth and Environments 15 (2): 185-210.

[8] Claudia Chwalisz. 2020. 'Reimagining Democratic Institutions: Why and How to Embed Public Deliberation.' In: OECD (org.) Innovative Citizen Participation and New Democratic Institutions: Catching the Deliberative Wave. OECD Publishing, Paris. https://doi.org/10.1787/339306da-en.

[9] COSLA. 2021. Community Choices: Participatory Budgeting in Scotland. Framework for the operation of the 1% target for Local Authorities (2021 Update). Available at: https://www.cosla.gov.uk/__data/assets/pdf_file/0017/26234/COSLA-SG-Participatory-Budgeting-Framework-Agreement-Jun e-2021.pdf

[10] Jonathan Davies, Miguel Arana-Catania, Rob Procter, Felix-Anselm van Lier, Yulan He. 2021. Evaluating the application of NLP tools in mainstream participatory budgeting processes in Scotland. 14th International Conference on Theory and Practice of Electronic Governance.

[11] Democratic Society. 2016. Digital tools and Scotland's Participatory Budgeting programme. The Democratic Society. Available at: https://www.demsoc.org/uploads/store/mediaupload/67/file/DS-Digital-Tools-paper.pdf

[12] John Dryzek, Andre Bächtiger, Simone Chambers, et al. 2019. The crisis of democracy and the science of deliberation. Science, 363 (6432): 1144–1146.

[13] Stephen Elstub and Oliver Escobar. 2019. Handbook of Democratic Innovation and Governance. Cheltenham, UK.

[14] Oliver Escobar, Fiona Garven, Chris Harkins, Kathleen Glazik, Simon Cameron, and Ali Stoddart. 2018. 'Participatory budgeting in Scotland: The interplay of public service reform, community empowerment and social justice.' In: N. Dias (org.) Hope for Democracy: 30 years of Participatory Budgeting Worldwide. Faro: Epopeia Records & Oficina.

[15] Elena Fagotto and Archon Fung. 2014. 'Embedding Public Deliberation in Community Governance.' In J. Girouard and C. Sirianni (Eds.) Varieties of Civic Innovation: Deliberative, Collaborative, Network, and Narrative Approaches. Vanderbilt University Press.

[16] Archon Fung and Erik Olin Wright. 2003. Deepening Democracy: Institutional Innovations in Empowered Participatory Governance. Verso, London.

[17] Greg Guest, Kathleen M. MacQueen and Emily E. Namey. 2012. Applied Thematic Analysis. London: Sage.

[18] Chris Harkins, Katie Moore, and Oliver Escobar. 2016. Review of 1st Generation Participatory Budgeting in Scotland. Edinburgh: What Works Scotland.

[19] Michelle E. Kiger and Lara Varpio.2020. Thematic analysis of qualitative data: AMEE Guide No. 131. Medical Teacher. DOI: 10.1080/0142159X.2020.1755030

[20] Penny Mackieson, Aron Shlonsky and Maria Connolly. 2018. Increasing rigor and reducing bias in qualitative research: A document analysis of parliamentary debates using applied thematic analysis. Qualitative Social Work, 18(6): 965-980

[21] Daniel Oross and Gabriella Kiss. 2021. More than just an experiment? Politicians arguments behind introducing participatory budgeting in Budapest. Acta Politica. https://doi.org/10.1057/s41269-021-00223-6

[22] Angela O'Hagan, Clementine Hill-O'Connor, Claire MacRae, Paul Teedon. 2019. Evaluation of Participatory Budgeting Activity in Scotland 2016-2018. Scottish Government.

[23] Michael Quinn Patton. 2002. Qualitative Evaluation and Research Methods (3rd edition). Thousand Oaks: Sage.

[24] Graham Smith. 2009. Democratic Innovations: Designing Institutions for Citizen Participation. Cambridge: Cambridge University Press.

[25] Graham Smith. 2019. 'Reflections on the theory and practice of democratic innovations'. In: S. Elstub and O. Escobar (Eds.) Handbook of Democratic Innovation and Governance. Cheltenham, UK.